\documentclass[12pt,preprint]{aastex}

\def\simlt{\mathrel{\spose{\lower 3pt\hbox{$\mathchar''218$}}
     \raise 2.0pt\hbox{$\mathchar''13C$}}}
\def\simgt{\mathrel{\spose{\lower 3pt\hbox{$\mathchar''218$}}
     \raise 2.0pt\hbox{$\mathchar''13E$}}}

\def\wig#1{\mathrel{\hbox{\hbox to 0pt{%
          \lower.5ex\hbox{$\sim$}\hss}\raise.4ex\hbox{$#1$}}}}

\def\v1n{{\cal U}^N_1}

\def\sss{\scriptscriptstyle}
\def\phih2h2{\phi_{\sss {\rm H_2-H_2}}}
\def\phh2{\phi_{\sss {\rm H-H_2}}}

\def\Teff{T_{\rm eff}}
\def\sqr#1#2{{\vcenter{\vbox{\hrule height.#2pt 
  \hbox{\vrule width.#2pt height#1pt \kern#1pt 
  \vrule width.#2pt} 
  \hrule height.#2pt}}}}

\def\ref#1{\parindent=0pt\hangindent24pt\hangafter1
           \baselineskip=20pt{#1}\par }

\begin{document}
\def\gtorder{\mathrel{\raise.3ex\hbox{$>$}\mkern-14mu
             \lower0.6ex\hbox{$\sim$}}}
\def\ltorder{\mathrel{\raise.3ex\hbox{$<$}\mkern-14mu
             \lower0.6ex\hbox{$\sim$}}}

\def\today{\number\year\space \ifcase\month\or  January\or February\or
        March\or April\or May\or June\or July\or August\or
        September\or
        October\or November\or December\fi\space \number\day}
\def\fraction#1/#2{\leavevmode\kern.1em
 \raise.5ex\hbox{\the\scriptfont0 #1}\kern-.1em
 /\kern-.15em\lower.25ex\hbox{\the\scriptfont0 #2}}
\def\spose#1{\hbox to 0pt{#1\hss}}
\def\heion{\ion{He}{2}}

\title{The Mid-Infrared Spectra of Brown Dwarfs}
\author{D. Saumon\footnote{Los Alamos National Laboratory, MS F699, Los Alamos, NM 87545}, 
M. S. Marley\footnote{NASA Ames Research Center, MS 245-5, Moffett Field, CA 94025}
\ and K. Lodders\footnote{Planetary Chemistry Laboratory, Department of Earth
and Planetary Sciences, Washington University, St. Louis, MO 63130-4899}}

\pagestyle{plain}

\begin{abstract}
We present an analysis of brown dwarf model spectra in the mid-infrared spectral
region (5 -- 20$\,\mu$m), in anticipation of data obtained with the Space Infrared Telescope
Facility.  The mid-infrared spectra of brown dwarfs are in several ways simpler than those in the 
near-infrared and yet provide powerful diagnostics of brown dwarf atmospheric physics and
chemistry, especially when combined with ground-based data.  We discuss the possibility of 
detection of new molecular species and of the silicate cloud,  predict strong observational
diagnostics for non-equilibrium chemistry between
CO and CH$_4$, and N$_2$ and NH$_3$, and speculate on the possibility of 
discovering brown dwarf stratospheres. 
\end{abstract}

\keywords{stars: low mass, brown dwarfs --- radiative transfer --- molecular processes
           --- infrared: stars}

\section{Introduction}

The near-infrared and optical spectra of brown dwarfs have received intense 
observational and theoretical
scrutiny (see the recent reviews by Basri 2000 and Burrows et al. 2001)  but their 
mid-infrared spectra 
have been relatively neglected both in observation and theory.  There are only two 
announced photometric detections
of a brown dwarf beyond $5\,\rm \mu m$ (Matthews et al. 1996; Creech-Eakman et al. 2003) 
and neither strongly constrains models. Although synthetic mid-IR spectra have been 
published (e.g. Marley et al. 1996; Tsuji, Ohnaka \& Aoki 1999; Allard et al. 2001; 
Burrows, Sudarsky \& Lunine 2003) there has yet been no detailed discussion of the expected
spectral diagnostics in L and T dwarfs.
With the Space Infrared Telescope Facility (SIRTF) now in orbit, the dramatic 
improvement in  our ability to
observe brown dwarfs in the mid-infrared will give new insights
in the astrophysics of the complex atmospheres of these cool, dim denizens of our neighborhood.
In this {\it Letter}, we present mid-IR model spectra of brown dwarfs and discuss their properties
in terms of effective temperature, important molecular absorbers, the role of silicate clouds,
and anticipate potential discoveries with SIRTF.

The instruments onboard SIRTF 
cover the wavelength range from 3 to 180$\,\mu$m.  Two of these will target 
brown dwarfs: the Infrared Array Camera (IRAC) and the Infrared Spectrograph (IRS).
Brown dwarfs will be imaged by IRAC in four bandpasses
centered at 3.6$\,\mu$m, 4.5$\,\mu$m, 5.8$\,\mu$m, and 8.0$\,\mu$m,
respectively (Fig. 1)\footnote{\tt http://sirtf.caltech.edu/SSC/irac}.  
Since brown dwarfs become very dim and their spectra contain little information beyond 20$\,\mu$m,
the most useful IRS observations will be in the ``Short wavelength, Low resolution'' 
(SL) and the less sensitive ``Short wavelength, High resolution'' (SH) 
modes\footnote{\tt http://sirtf.caltech.edu/SSC/irs}.  
The SL and SH modes cover the 5.3 -- 14.2$\,\mu$m  and the 10.0 -- 19.5$\,\mu$m 
spectral bands, respectively.

\section{Atmosphere models and spectra}

To model the atmospheres and spectra of the L- and T-dwarfs we employ the
radiative-convective equilibrium atmosphere model of Marley et
al. (1996; further described in Burrows et al. 1997 and Marley et al. 2002) which
includes the precipitating cloud model of Ackerman \&
Marley (2001). High resolution spectra are computed from the
initial temperature profile and cloud structure (Saumon et al.
2000; Geballe et al. 2001).  The chemistry is computed in the
framework of the cloud condensation model
(Lodders 1999a; Lodders \& Fegley 2002; Lodders 2002; Lewis 1969).  The opacity 
includes the molecular lines
of H$_2$O, CH$_4$, CO, NH$_3$, H$_2$S, PH$_3$, TiO, VO, CrH, FeH, CO$_2$,
HCN, C$_2$H$_2$, C$_2$H$_4$, C$_2$H$_6$
complemented
with the atomic lines of the alkali metals (Li, Na, K, Rb and Cs) and continuum
opacity sources from H$_2$ CIA, H$_2$, H and He Rayleigh scattering, H$^-$ bf and ff,
H$_2^-$ ff, He$^-$ ff, and H$_2^+$ bf and ff.  While the chemical equilibrium is
computed with a large number of condensates, only Fe, MgSiO$_3$, Al$_2$O$_3$,
$\rm H_2O$, and $\rm NH_3$ are considered in the cloud model. Condensed $\rm Mg_2SiO_4$ is
accounted for with MgSiO$_3$ and the remaining condensates are not appreciable sources
of opacity.

We consider spectra from $\Teff=600$ to 2400$\,$K which covers
the full range of L and T dwarfs.  We limit the discussion to models with
solar metallicity.  The sedimentation parameter of the cloud
model is fixed at $f_{\rm sed}=3$, which gives a good representation of far-red
and near-IR photometry of L dwarfs (Marley et al. 2002; Burgasser et al. 2002)  
and of the ammonia cloud deck of Jupiter as well (Ackerman \& Marley 2001).
For simplicity, we do not take into account the possibility of cloud disruption
near the L/T boundary (Burgasser et al. 2002).  

\section{Mid-infrared spectra of brown dwarfs}

The combination of IRAC photometry and IRS spectroscopy with
ground-based optical and near-IR data will give the complete spectral energy 
distributions (SED) of many brown dwarfs.  Empirical bolometric
corrections immediately follow as well as the bolometric luminosity for objects 
with known parallaxes.  An independent determination of $\Teff$ (e.g. by fitting spectra
with models) will give the gravity, radius, and mass of individual brown dwarfs.

It is well known that in the far-red and near-IR, the brightness temperature  $T_{\rm br}$ 
of brown dwarfs varies strongly with wavelength due to the high contrast between opacity
windows and molecular absorption bands.  This is still true in the mid-IR
up to about 11$\,\mu$m, thereafter $T_{\rm br}$ gradually decreases
to stabilize at $\sim$75\% of $\Teff$ beyond 18$\,\mu$m.  Only near 6$\,\mu$m and
near 10$\,\mu$m does  $T_{\rm br}$ become as large as $\Teff$.  
For $4 \wig< \log g \,({\rm cgs}) \wig< 5.5$,
the pressure at the mid-IR photosphere stays between 0.1 and 3$\,$bar, depending mostly
on wavelength and surface gravity and least on $\Teff$.  Basically,
the mid-IR spectra of brown dwarfs are formed at $T\wig< \Teff$
and pressures of about 1$\,$bar.

\subsection{Molecular species}

Figure 1 identifies the mid-IR molecular absorbers in a sequence of 
spectra from 600 to 2400$\,$K.   By far, the most important absorber is H$_2$O
throughout the entire spectral range, except for two strong molecular bands due
to CH$_4$ and NH$_3$.  Beyond 15$\,\mu$m,  all features are due to H$_2$O, with the
exception of a few weak NH$_3$ bands between 32 and 46$\,\mu$m for models with
 $\Teff \wig< 800\,$K.
In the high-$\Teff$ spectra, the fundamental band of CO at 4.8$\,\mu$m and a weak
TiO band at 10$\,\mu$m are the only features not originating from H$_2$O.  The step in
the spectrum $\sim 6.5\,\mu$m is a H$_2$O feature.  The TiO band disappears
at 2200$\,$K but the CO band persists down to 900$\,$K in these cloudy models.
As $\Teff$ decreases, new molecular bands appear and steadily increase in strength.
The 10.5$\,\mu$m band of NH$_3$ appears at 1150$\,$K. 
This band is very broad and dominates the spectrum from 8.5 to 16$\,\mu$m
at $\Teff \le 800\,$K. Weaker bands of NH$_3$ appear at 900$\,$K (5.5 -- 7$\,\mu$m)
and 800$\,$K (3.9 -- 4.5$\,\mu$m).  The former is superimposed
on a strong H$_2$O band, however. Methane shows only one band in the mid-IR,
centered at 7.8$\,\mu$m.  It appears at 1600$\,$K and becomes very strong at
low $\Teff$ to dominate the spectrum between 7 and 9.2$\,\mu$m. 
Broad, featureless CIA opacity appears faintly in the 9 to 13$\,\mu$m region in cool
models below 1000$\,$K, but it shows no gravity sensitivity and is probably
undetectable.

Our current phosphorus chemistry uses updates for several P-gases (PH, PH$_3$,
PN, PS; Lodders 1999b, 2004) and the P$_4$O$_6$ gas thermodynamic properties from
Gurvich, Veyts \& Alcock (1989) instead of those from Chase (1998). Our previous
models used P$_4$O$_6$ data from Chase (1998) and calculated spectra did not show
the strongest PH$_3$ band centered at 4.3$\,\mu$m, just to the blue of the
fundamental CO band. However, the Gurvich et al. data make P$_4$O$_6$
significantly less stable which results in a larger abundance of PH$_3$ and
the band appears below 1100$\,$K with a characteristic narrow absorption
spike.  This feature should provide a good observational test of the phosphorus chemistry
in brown dwarfs, but its signature in IRAC band 2 photometry will have to be distinguished from
possibly enhanced CO (Section 5).

The most abundant sulfur-bearing compound is H$_2$S, a molecule that has 
yet to be detected in brown dwarfs.  Mid-IR observations will not improve its prospects of being
noticed since its opacity is consistently
3 -- 4 orders of magnitude below that of H$_2$O.  

We predict that CO and CH$_4$ will be visible simultaneously in the mid-IR for 
$\Teff \wig<1600\,$K, i.e. from mid-L to mid-T spectral types.  The
NH$_3$ band at 10.5$\,\mu$m will be easy to detect and is a characteristic of all
T dwarfs (see Section 5, however).  Observations of Gl 229B should 
confirm the weak detection of this molecule in the near-IR (Saumon et al. 2000).

Figure 1 shows that the mid-IR spectra of brown dwarfs  are characterized by
molecular bands with strong $\Teff$ dependences below 1600$\,$K.
Fitting the IRS spectra alone with models should allow the determination of $\Teff$ 
of brown dwarfs of mid-L and later spectral types.

\subsection{Silicate cloud}

The effects of silicate, iron, and corundum clouds on the near-IR spectra of brown
dwarfs have been discussed extensively (Ackerman \& Marley 2001; Allard et al. 2001;
Burgasser et al. 2002; Cooper et al. 2003; Marley et al. 2002; 
Tsuji 2002; Tsuji \& Nakajima 2003).  Because the mid-IR photosphere is characterized by a fairly
constant brightness temperature and a single dominant absorber (H$_2$O),
the spectral signature of clouds stands out more clearly than in the near-IR.
Most interesting is the possibility of detecting the silicate cloud  opacity feature 
$\sim 10\,\mu$m.  Figure 2 shows spectra  computed with and without the cloud
opacity using the same cloudy atmospheric structure.   We find that the silicate
feature (a rise in the opacity between 9 and 10$\,\mu$m to a new plateau)
is masked by a similar behavior in the H$_2$O and CH$_4$ opacities.
Nevertheless, the cloud opacity is large enough to appear 
in two windows where the gas is less opaque in models with $1200 \wig< \Teff
\wig< 2000\,$K: from $\sim 4$ to 6.5$\,\mu$m (affecting IRAC bands 1 through 3) 
and from 9 to 13$\,\mu$m.  The effect is strongest at $\Teff \sim 1600\,$K, which 
corresponds to the largest cloud optical depth above the 10$\,\mu$m
photosphere.  Because the silicate absorption flattens the spectrum
noticeably in the 9 to 12$\,\mu$m range, its identification should be
fairly straightforward in IRS spectra of mid-L dwarfs.  For models with a higher
condensate sedimentation efficiency ($f_{\rm sed}=5$), the thinner cloud layer remains
optically thin above the photosphere and is invisible at 10$\,\mu$m. 
This feature can thus
serve as a diagnostic of the vertical structure of the cloud model as its
strength depends strongly on the cloud's vertical extent and on the particle
size. 

\subsection{IRAC colors of brown dwarfs}

The IRAC will image brown dwarfs and provide photometry in 4 broad bandpasses 
(Fig. 1), labeled Band 1 through
4.  Bands 1 and 2 cover wavelengths not accessible to the IRS.  

As can be seen in Fig. 1, Band 2 covers the bands of CO and PH$_3$ 
but is nevertheless the least affected by strong molecular bands
over the full range of $\Teff$ of interest.  Bands 1 and 4 are increasingly
affected by CH$_4$ absorption below $\Teff \sim 1500\,$K.  Strong H$_2$O absorption  
appears in Band 3 below $\Teff \sim 1200\,$K.

A study of color-color diagrams
shows that IRAC colors will be most useful as diagnostics of brown 
dwarf physical properties for $\Teff \wig< 1400\,$K.  At higher $\Teff$,
the colors change little with $\Teff$ or the $\Teff$ and gravity dependences become
hopelessly tangled.  For brown dwarfs with known distances, color-magnitude diagrams   
provide a good discriminant of both $\Teff$ and gravity.  

\section{Trace molecular species}

The IR spectra of giant planets reveal the presence of trace carbon species such as
CO$_2$, C$_2$H$_2$, and HCN in {\it emission}.  These arise primarily
from the photochemistry of CH$_4$ (also seen in emission)  in the stratosphere, i.e. above a 
temperature inversion in the atmosphere (Moses 2000).  No published  brown dwarf model
shows a temperature inversion, mainly because
the present knowledge of the physics of brown dwarf atmospheres is 
too primitive to model a credible stratosphere.  We nevertheless 
anticipate that brown dwarfs also may have stratospheres (Yelle 2000)
and observations in the mid-IR
are a powerful way to discover them by detecting emission features from trace
molecules such as CO$_2$, C$_2$H$_2$, C$_2$H$_4$, C$_2$H$_6$, and HCN that should
arise from photochemistry driven by the stratospheric and possibly the weak
interstellar UV flux.  In L and T dwarfs,
H$_2$O remains in the gas phase and can participate in the photochemistry, 
leading to the formation of species such as HCO and CH$_2$O (Friedson, Wilson \& Moses 2003). 

In the absence of a stratosphere, these molecules would appear in {\it absorption}.
The equilibrium abundances, i.e. in the absence of photochemistry, are all
quite low.  We have calculated the minimum enhancement factor ($\epsilon$)
 of the abundance of several trace species required for detection. We 
define $\epsilon$  as the ratio of the column
density of the trace species above the photosphere that is needed for detection to
the equilibrium column density above the photosphere.  For each species in Table 1,
we give the wavelength of the strongest opacity feature, the $\Teff$ that is
most favorable for detection (i.e. giving the lowest enhancement factor), and
the enhancement factor.
Clearly, none of these species is likely to be detected in absorption, except
for CO$_2$, which would be favored by higher metallicity and lower gravity.
For instance, the equilibrium abundance of CO$_2$ should be marginally
detectable at 15$\,\mu$m in a model with $\Teff=1200\,$K and $\log g=4$.

Objects with masses below 13$\,M_J$ do not burn deuterium and deuterated molecular
species are expected in their atmospheres.  With its distinctive 4.55$\,\mu$m band 
and a relatively large abundance, CH$_3$D is the most easily detected deuterated 
species expected in brown dwarfs.  We find that for CH$_3$D/CH$_4=2 \times 10^{-5}$,
an enhancement factor of $\wig> 250$ would be necessary to detect it in T dwarfs
above $\Teff=600\,$K (Table 1).  Detection becomes easier at lower $\Teff$ but is
very unlikely among the known objects.
  
\section{Non-equilibrium chemistry}

Vertical transport in brown dwarf atmospheres can lead to non-equilibrium
abundances in slowly reacting species such as CO and N$_2$. 
The net effect is to enhance the upper atmosphere abundances of CO and N$_2$ 
to the detriment of CH$_4$, H$_2$O and NH$_3$
(Fegley \& Lodders 1996; Griffith \& Yelle 1999; Lodders \& Fegley 2002).
The detection of greatly enhanced CO (Noll, et al. 1997) and
of depleted NH$_3$ (Saumon et al. 2000) in Gl 229B, as well as the $M^\prime$
flux deficiency observed in several T dwarfs (Golimowski et al. 2004) suggest that 
non-equilibrium chemistry is a general phenomenon in brown dwarfs.  Nevertheless,
it remains mostly unexplored observationally.

We have modeled the non-equilibrium spectra of brown dwarfs following the method
described in Saumon et al (2003).  We find that the 7.8$\,\mu$m band of CH$_4$ 
appears at $\Teff \sim 1400\,$K and 1200$\,$K for
mixing coefficients of $K_{zz}=10^2$ and $K_{zz}=10^4\,$cm$^2$/s, respectively, 
compared to 1600$\,$K in the case of equilibrium chemistry.
The 10.5$\,\mu$m band of NH$_3$ appears at $\Teff \wig< 1000\,$K for 
$K_{zz}\wig> 10^2\,$cm$^2$/s, instead of 1150$\,$K.  Because of the the relatively 
flat equilibrium abundance profile of NH$_3$, this result is nearly independent of
$K_{zz}$.  While NH$_3$ can be depleted by more than one order of magnitude
(for $\Teff=800\,$K), the 10-11$\,\mu$m features remain strong (Fig. 3).

\section{Conclusions}

Spectroscopy with the IRS instrument on SIRTF between 5 and 20$\,\mu$m will complete the
sampling of the spectral energy distribution of brown dwarfs.
From our analysis of our mid-IR synthetic spectra, we find that the new data will:
1) clearly reveal the presence of NH$_3$ in T dwarfs, despite the strong
   depletion due to vertical transport in the atmosphere,
2) likely detect the silicate cloud $\sim 10\,\mu$m in mid-L dwarfs and
put strong constraints on the vertical structure of the cloud, 
3) lead to a much more complete picture of non-equilibrium
chemistry among CO, CH$_4$, H$_2$O, N$_2$ and NH$_3$, 
4) show CO$_2$ in low-gravity, high metallicity targets near the L/T transition,
5) {\it not} reveal interesting species such as H$_2$S, CH$_3$D, and the
   H$_2$ CIA opacity, and
6) possibly discover brown dwarf stratospheres through
 emission from trace species such as CO$_2$, HCN, HCO, C$_2$H$_2$, C$_2$H$_4$, 
C$_2$H$_6$ and CH$_2$O.
Furthermore, we predict that: 7) a strong band of PH$_3$  at 4.3$\,\mu$m
falls in the IRAC band 2, 8) IRAC photometry will be most useful for
objects below 1400$\,$K, and 9) photospheric spectra of brown dwarfs
are of very little interest beyond 20$\,\mu$m.
The mid-IR spectral window is rich in diagnostics of 
brown dwarf atmospheres.  Our understanding of these cool neighbors
will grow dramatically with our first mid-IR observations with SIRTF.

\acknowledgments 
We thank Richard Freedman for providing the molecular 
opacities used in this calculation and the SIRTF/IRS instrument team 
for informative discussions. 
This work was supported in part by NASA grants NAG2-6007 and NAG5-8919 and NSF 
grant AST 00-86288 (MSM) and by NSF grant AST 00-86487 (KL). Part of this work was 
supported by the United States Department of Energy under contract W-7405-ENG-36

\clearpage

\figcaption[fig1.ps]{Sequence of mid-IR spectra of brown dwarfs with $\Teff$ ranging
from 600 to 2400$\,$K (from bottom to top) in steps of 200$\,$K. All 
models have solar metallicity, $\log g=5$, $f_{\rm sed}=3$ and are shown 
at a spectral resolution  of 200.  Spectra are shifted vertically for 
clarity.  Molecular bands  that appear at high $\Teff$ are
identified at the top, those that appear at low $\Teff$ are shown at 
the bottom.  Outside of those bands, H$_2$O is responsible for all features.
Weak bands are indicated with dashed lines.
The bandpasses of the four IRAC filters are shown by thick solid lines
at the top.
\label{fig:sp_sequence}}

\figcaption[fig2.ps]{Effect of silicate cloud opacity on the mid-IR spectra of
brown dwarfs.  The spectra shown have $\Teff=1000$ to 2400$\,$K in
steps of 200$\,$K.  The dotted lines show spectra
computed from the same cloudy $(T,P)$ structures and chemistry,
but without the cloud
opacity. IRAC bands 3 and 4 are shown by thick solid lines at the top.
\label{fig:SiO3_cloud}}

\figcaption[fig3.ps]{Effect of non-equilibrium chemistry on the mid-IR 
NH$_3$ band.  The spectra shown have $\Teff=800$ to 1600$\,$K in
steps of 200$\,$K.  The spectra shown are computed
computed from the same cloudy $(T,P)$ structures but with equilibrium
(solid) and non-equilibrium (dotted) chemistry. See Fig. 2 for details.
\label{fig:NH3_noneq}}

\begin{figure}
\plotone{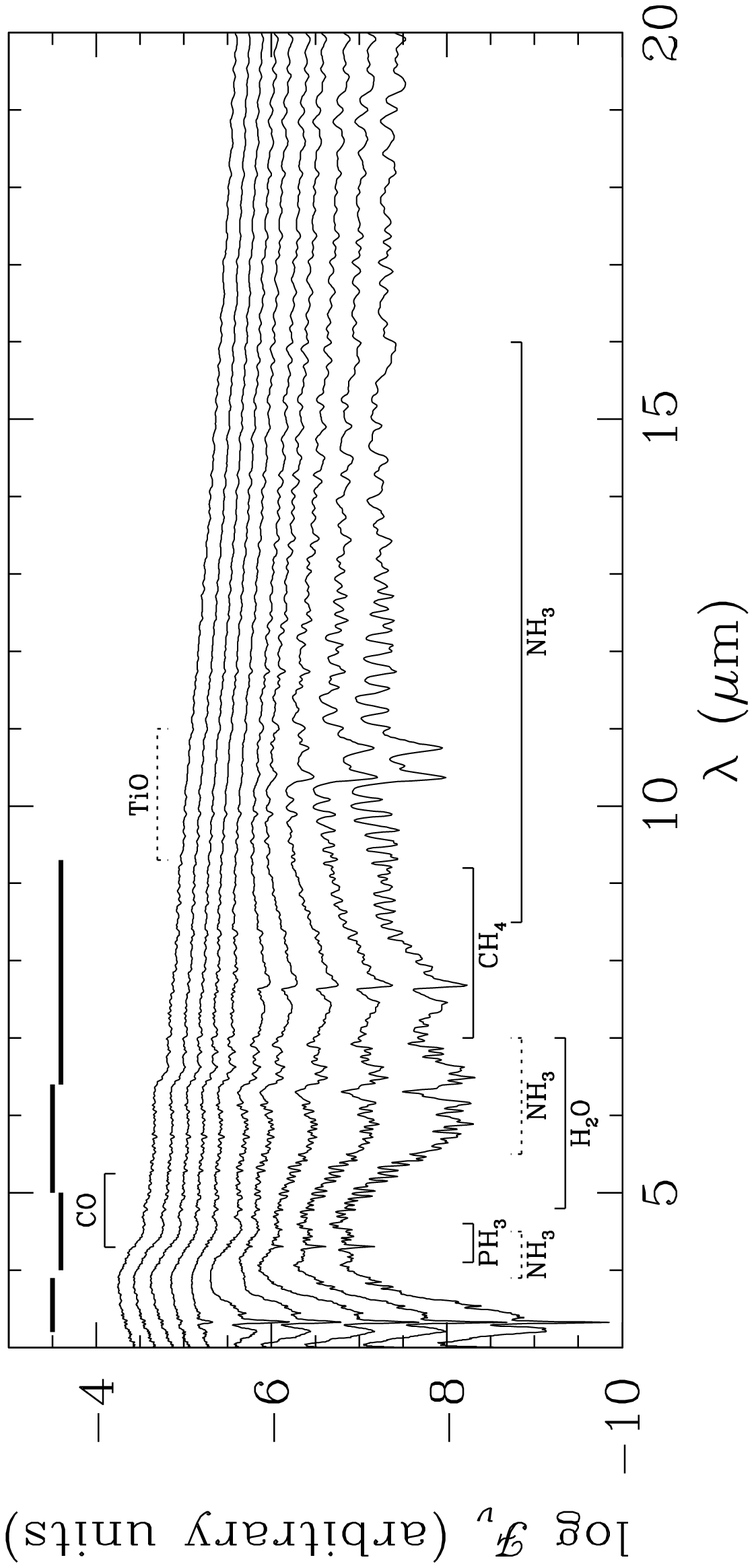}
\end{figure}

\begin{figure}
\plotone{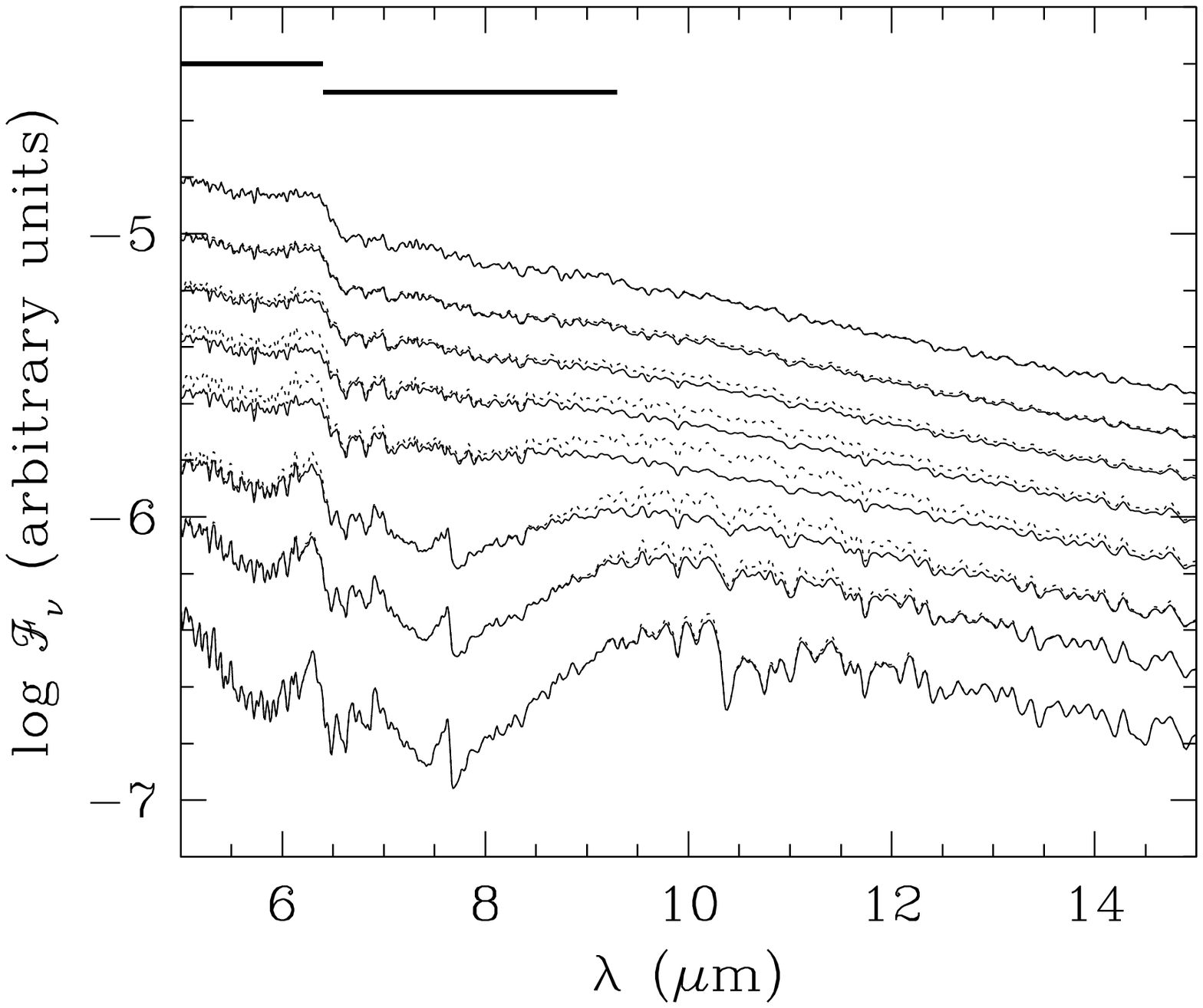}
\end{figure}

\begin{figure}
\plotone{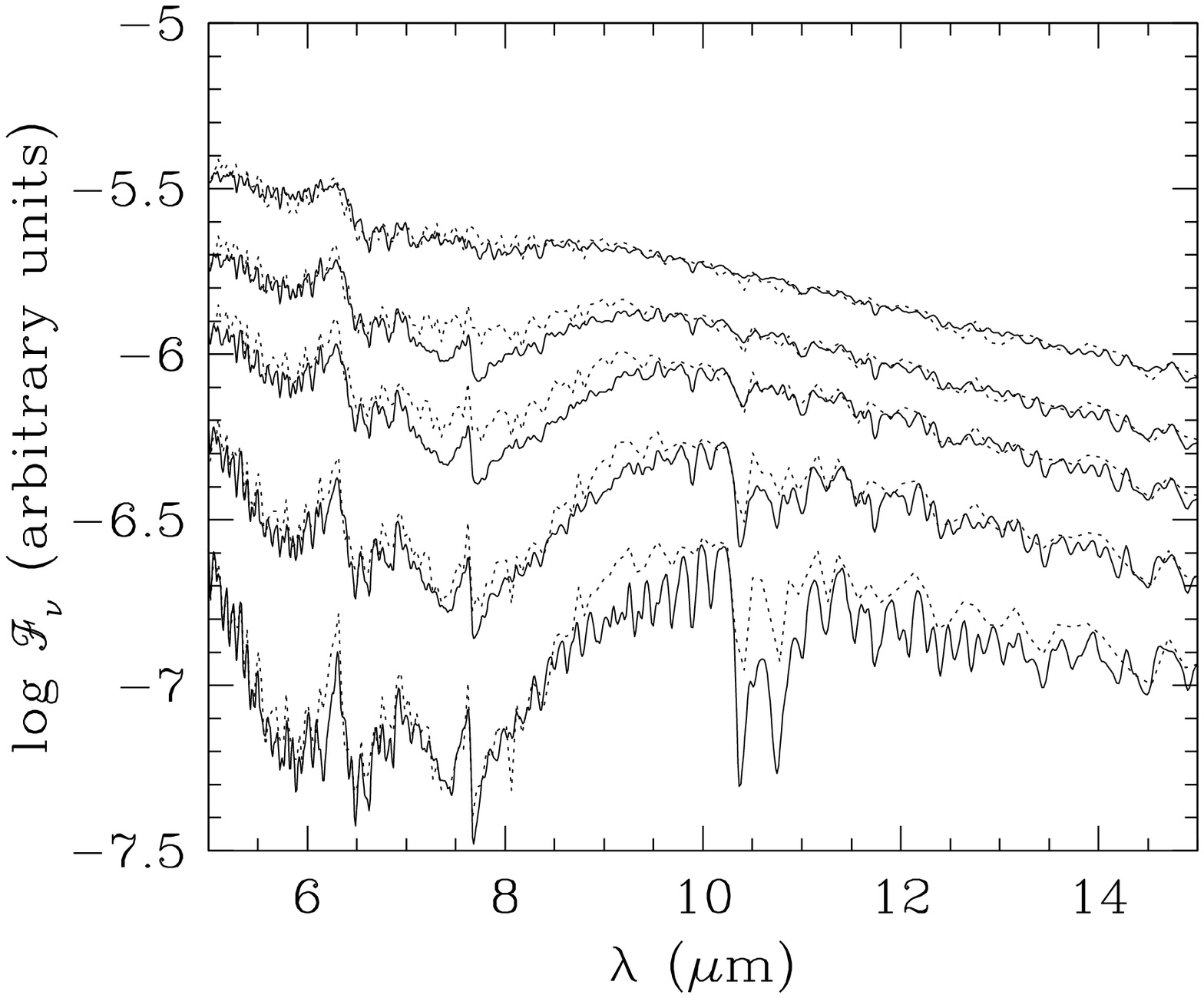}
\end{figure}

\clearpage

\begin{deluxetable}{cccc}
\tablecaption{Enhancement factors for trace species.\tablenotemark{a}}
\tablewidth{0pt}
\tablehead{
\colhead{Species} & $\lambda (\mu{\rm m})$ & \colhead{$\Teff$ (K)}   & \colhead{$\log(\epsilon)$}
}
\startdata
CO$_2$     & 15.0  & 1400  & 0.7 \\
HCN        & 14.0  & 1400  & 2.7 \\
C$_2$H$_2$ & 13.7  & 1400  & 5.8 \\
C$_2$H$_4$ & 10.5  & 1200  & 4.9 \\
C$_2$H$_6$ & $\sim 12.2$ &1000  & 3.5 \\
CH$_3$D    & 4.55  & $<600$  & $<2.4$\tablenotemark{b} \\
\enddata

\tablenotetext{a}{For $\log g=5$ and solar metallicity.}
\tablenotetext{b}{Cloudless atmosphere model.}
\end{deluxetable}

\clearpage
\end{document}